\documentstyle{article}

\def\a{\alpha}
\def\d{\delta}
\def\lb{\lambda}
\def\om{\omega}
\def\s{\sigma}
\def\t{\theta}
\def\ve{\varepsilon}
\def\vp{\varphi}

\def\D{\Delta}
\def\Lb{\Lambda}
\def\Om{\Omega}
\def\Si{\Sigma}
\def\T{\Theta}

\def\bsh{\backslash}
\def\fl{\forall}
\def\ify{\infty}
\def\mpo{\mapsto}
\def\nb{\nabla}
\def\op{\oplus}
\def\ot{\otimes}
\def\sbs{\subset}
\def\ts{\times}
\def\wdg{\wedge}

\def\ra{\rightarrow}
\def\longra{\longrightarrow}

\def\Ec{{\cal E}}
\def\Fc{{\cal F}}
\def\Sc{{\cal S}}

\font\tenbb=msbm10
\font\sevenbb=msbm7
\font\fivebb=msbm5
\newfam\bbfam
\textfont\bbfam=\tenbb 
\scriptfont\bbfam=\sevenbb
\scriptscriptfont\bbfam=\fivebb
\def\bb{\fam\bbfam}

\def\Cb{{\bb C}}
\def\Nb{{\bb N}}
\def\Qb{{\bb Q}}
\def\Rb{{\bb R}}
\def\Tb{{\bb T}}
\def\Zb{{\bb Z}}

\newtheorem{theorem}{Theorem}
\newtheorem{lemma}[theorem]{Lemma}
\newtheorem{definition}[theorem]{Definition}
\newtheorem{proposition}[theorem]{Proposition}

\def\build#1_#2^#3{\mathrel{
\mathop{\kern 0pt#1}\limits_{#2}^{#3}}}

\title{$C^*$  {\bf algebras and Differential Geometry
}\footnote{This paper appeared in March 1980
as {\sc A. Connes}, {\it  $C^*$ alg\`ebres et
g\'eom\'etrie differentielle}.  C.R. Acad. Sci. Paris,
Ser.~A-B , 290, 1980.}}
\author{Alain Connes} 
\date{}

\begin{document}

\maketitle

\abstract{

This is the translation to appear in the "SUPERSYMMETRY 2000 - Encyclopaedic Dictionary"
 of the original Comptes Rendus Note, 
published in March 1980, in which basic notions of noncommutative geometry
were introduced and applied to noncommutative tori. These include connections
on finite projective modules, their curvature, and the Chern character.
Finite projective modules on the noncommutative two-torus $ \Tb^2_{\theta}$
were realized as Schwartz spaces of vector valued functions on $\Rb$.
Explicit constant curvature connections were constructed and a basic integrality
phenomenon of the total curvature was displayed. 
The pseudo-differential calculus and the Atiyah-Singer
index theorems were extended to Lie group actions on $C^*$ algebras and 
used to explain the above integrality of the total curvature by an index formula
for finite difference-differential operators on the line.
Recent interest in the hep-th literature for basic notions of 
noncommutative geometry in the case of noncommutative tori 
(cf for instance hep-th/0012145 for an excellent review)
prompted us to make the English translation of the original paper available.

}

\vglue 1cm

\section*{Introduction}

The theory of $C^*$ algebras is a natural extension of the topology 
of locally compact spaces. A large portion of the theory has been dealing
with the analogue of Radon measures.
Among the invariants of algebraic topology it is $K$ theory
which is easiest to adapt to the framework of $C^*$ algebras. 
The study of the analogue of a differentiable structure on the 
locally compact space $X$ was proposed in \cite{6} as the 
investigation of derivations of a $C^*$ algebra $A$.
In this note we shall develop the basic notions of differential topology
in the special case where the differential structure is obtained through
a Lie group $G$ of automorphisms of $A$. 
We extend the notions of connection on a bundle, of curvature 
and Chern classes and construct, given a  
$G$-invariant trace $\tau$, a morphism ${\rm 
ch}_{\tau}$ from $K_0 (A)$ to $H_{\Rb}^* (G)$, (the cohomology of 
left invariant differential forms on $G$).
We then extend the Atiyah-Singer index theorem.
All this study is motivated by a simple example, that of the 
irrational rotation $C^*$ 
algebra $A_{\t}$ of angle $\t$. 
This algebra is highly noncommutative, being simple and non type 
I, and Pimsner and Voiculescu computed $K(A_{\t})$ \cite{3}.
 We first endow $A_{\t}$ with the differential structure coming from
the obvious action of the compact group  $\Tb^2$, $\Tb = \{ z \in \Cb , 
\, \vert z \vert = 1 \}$, we then show that the space of smooth ($C^{\ify}$)
sections of the vector bundle of dimension $\t$ on 
$A_{\t}$ is identical with the Schwartz space $\Sc (\Rb)$, on which 
 $A_{\t}$ acts by finite difference operators. 
The following operators
$$
D_1 \, f = \frac{d}{dt} \ f \qquad \hbox{and} \qquad D_2 \, f = 
\frac{2 \, \pi \, it}{\t} \ f 
$$
then define a connection (of constant curvature $1/\t$) and
 the index theorem allows to compute the index (integer valued)
of polynomials in $D_1$, $D_2$ with finite difference operators
as coefficients.

\smallskip

Let $(A,G,\a)$ be a $C^*$ dynamical system, where $G$ is a Lie group. 
We shall say that $x \in A$ is of $C^{\ify}$ class iff 
the map $g \mpo \a_g (x)$ from $G$ to the normed space $A$ is 
 $C^{\ify}$. The involutive algebra $A^{\ify} = \{ x \in A , 
\, x \ \hbox{of class} \ C^{\ify} \}$ is norm dense in $A$.

Let $\Xi^{\ify}$ be a finite projective module on $A^{\ify}$,
 (we shall write it as a right module) ~; $\Xi^0 = \Xi^{\ify} 
\ot_{A^{\ify}}$ $A$ is then  a finite projective module on
$A$.

\medskip

\begin{lemma}
For every finite projective module $\Xi$ on $A$, there exists a 
finite projective module $\Xi^{\ify}$ on $A^{\ify}$, unique up to
isomorphism, such that $\Xi$ is isomorphic to $\Xi^{\ify} 
\ot_{A^{\ify}} \, A$.
\end{lemma}

\medskip

In the sequel we let $\Xi^{\ify}$ be a finite projective module 
on $A^{\ify}$. An hermitian structure on 
$\Xi^{\ify}$ is given by a positive hermitian form  $\langle 
\xi , \eta \rangle \in A^{\ify}$, $\fl \, \xi , \eta \in \Xi^{\ify}$ 
such that
$$
\langle \xi \cdot x , \eta \cdot y \rangle = y^* \langle \xi , \eta 
\rangle \, x \ , \qquad \fl \, \xi , \eta \in \Xi^{\ify} \, , \quad \fl \, 
x,y \in A^{\ify} \, .
$$
For $n \in \Nb$, $\Xi^{\ify} \ot \Cb^n$ is a finite projective module
on $M_n (A^{\ify}) = A^{\ify} \ot M_n (\Cb)$, this allows, 
replacing $A$ by $M_n (A) = A \ot M_n (\Cb)$ (and the 
$G$-action by $\a \ot {\rm id}$) to assume the existence of a  
selfadjoint idempotent $e \in A^{\ify}$ and of an isomorphism $F$ with the module 
$e \, A^{\ify}$ on $\Xi^{\ify}$. We then endow $\Xi^{\ify}$ with
the following hermitian structure~:
$$
\langle \xi , \eta \rangle = F^{-1} (\eta)^* \, F^{-1} (\xi) \in 
A^{\ify} \, .
$$
Let $\d$ be the representation of Lie$G$ in the Lie-algebra of derivations of
 $A^{\ify}$ given by
$$
\d_X (x) = \lim_{t \ra 0} \ \frac{1}{t} \, (\a_{g_t} (x) - x) \, , 
\qquad \hbox{where} \quad \dot{g}_0 = X \, , \ x \in A^{\ify} \, .
$$

\medskip

\begin{definition}
Let $\Xi^{\ify}$ be a finite projective module on $A^{\ify}$,  
a connection (on $\Xi^{\ify}$) is a linear map 
$\nb$ of $\Xi^{\ify}$ in $\Xi^{\ify} \ot ({\rm Lie} \ G)^*$ such that, 
for all $X \in {\rm Lie} \ G$ and $\xi \in \Xi^{\ify}$, $x \in 
A^{\ify}$ one has
$$
\nb_X (\xi \cdot x) = \nb_X (\xi) \cdot x + \xi \cdot \d_X (x) \, .
$$
\end{definition}

\medskip

We shall say that $\nb$ is compatible with the hermitian structure 
iff~:
$$
\langle \nb_X \, \xi , \xi' \rangle + \langle \xi , \nb_X \, \xi' 
\rangle = \d_X \langle \xi , \xi' \rangle \, , \qquad \fl \, \xi , 
\xi' \in \Xi^{\ify} \, , \quad \fl \, X \in \hbox{Lie} \ G \, .
$$
Every finite projective module, $\Xi^{\ify}$, on $A^{\ify}$ 
admits a connection~; on the module $e \, A^{\ify}$ the following 
formula defines the  {\it Grassmannian connection }
$$
\nb_X^0 (\xi) = e \, \d_X (\xi) \in e \, A^{\ify} \, , \qquad \fl \, \xi 
\in e \, A^{\ify} \, , \quad \fl \, X \in \hbox{Lie} \ G \, .
$$
This connection is compatible with the hermitian structure
$$
\langle \xi , \eta \rangle = \eta^* \, \xi \in A^{\ify} \, , \qquad 
\fl \, \xi , \eta \in e \, A^{\ify} \, .
$$

To the representation $\d$ of Lie $G$ in the Lie-algebra of derivations of
 $A^{\ify}$ corresponds the complex $\Om = A^{\ify} \ot 
\Lb \ (\hbox{Lie} \ G)^*$ of left-invariant differential forms on $G$ 
with coefficients in $A^{\ify}$. We endow $\Om$ with the algebra 
structure given by the tensor product of $A^{\ify}$ by the exterior algebra of 
 $(\hbox{Lie} \ G)_{\Cb}^*$
(we use the notation $\om_1 \wdg \om_2$ for the product 
of $\om_1$ with $\om_2$, one no longer has, of course, 
the equality
$\om_2 \wdg \om_1 = (-1)^{\partial \om_1 \, \partial \om_2} \, \om_1 
\wdg \om_2$).
The exterior differential
$d$ 
is such that~:
\begin{itemize}
\item[1$^{\rm o}$] for $a \in A^{\ify}$ and $X \in \hbox{Lie} \ G$ one 
has $\langle X , da \rangle = \d_X (a)$~;
\item[2$^{\rm o}$] $d (\om_1 \wdg \om_2) = d \om_1 \wdg \om_2 + 
(-1)^p \, \om_1 \wdg d \om_2$, $\fl \, \om_1 \in \Om^p$, $\fl \, \om_2 
\in \Om$~;
\item[3$^{\rm o}$] $d^2 \, \om = 0$, $\fl \, \om \in \Om$.
\end{itemize}
As $A^{\ify} \sbs \Om$, $\Om$ is a bimodule on $A^{\ify}$.

Every connection on $e \, A^{\ify}$ is of the form $\nb_X (\xi) = 
\nb_X^0 (\xi) + \t_X \, \xi$, $\fl \, \xi \in e \, A^{\ify}$, $X \in 
\hbox{Lie} \ G$, where the form $\t \in e \, \Om^1 \, e$ is uniquely 
determined by $\nb$, one has $\t_X^* = -\t_X$, $\fl \, X \in 
\hbox{Lie} 
\ G$ iff $\nb$ is compatible with the hermitian structure of $e \, 
A^{\ify}$.

\medskip

\begin{definition}
Let $\nb$ be a connection on the finite projective module 
$\Xi^{\ify}$ (on $A^{\ify}$), the curvature of $\nb$ 
is the element
$\T$ of ${\rm End}_{A^{\ify}} (\Xi^{\ify}) \ot \Lb^2 ({\rm Lie} \ 
G)^*$ given by
$$
\T (X,Y) = \nb_X \nb_Y - \nb_Y \nb_X - \nb_{[X,Y]} \in {\rm 
End}_{A^{\ify}} (\Xi^{\ify}) \, , \qquad \fl \, X,Y \in {\rm Lie} \ G 
\, .
$$
\end{definition}

\medskip

We identify ${\rm End} (e \, A^{\ify})$ with $e \, A^{\ify} \, e \sbs 
A^{\ify}$, the curvature $\T_0$ of the grassmannian connection is
the 2-form $e(de \wdg de) \in \Om^2$, the curvature of $\nb = \nb^0 + 
\t \wdg$ is equal to $\T_0 + e (d\t + \t \wdg \t) \, e \in \Om^2$.

\medskip

\begin{lemma}
With the above notations one has
$$
e (d\T) \, e = \T \wdg \t - \t \wdg \T \, .
$$
\end{lemma}

\medskip

Let now $\tau$ be a finite $G$-invariant trace on $A$. For  all
$k \in \Nb$, there exists a unique $k$-linear map from 
$\Om \ts \underbrace{\ldots}_{k \, {\rm times}} \ts \, \Om$ to $\Lb 
(\hbox{Lie} \ G)_{\Cb}^*$ such that,
$$
\tau^k(a_1 \ot \om_1, ... ,a_k \ot \om_k) = \tau(a_1 a_2...a_k) \,  \om_1
\wdg ... \wdg \om_k
$$
\medskip

\begin{proposition}
With the above notations the following invariant differential form on $G$, 
 $\tau^k (\T , \ldots , \T) \in \Lb^{2k} (\hbox{Lie} \ 
G)_{\Cb}^*$ is closed and its cohomology class only depends on 
the finite  projective module $\Xi^{\ify} = e \, A^{\ify}$ on 
$A^{\ify}$.
\end{proposition}

\medskip

We let $H_{\Rb}^* (G)$ be the cohomology ring of left invariant 
differential forms on  $G$, for $e$ as above we let
$$
{\rm Ch}_{\tau} (e) = \hbox{class of} \ \sum \left( \frac{1}{2 \, \pi 
\, i} \right)^k \ \frac{1}{k!} \ \tau^k (\T , \ldots , \T) \in 
H_{\Rb}^* (G) \, .
$$
One thus obtains a morphism ${\rm ch}_{\tau}$ from the group $K_0 (A)$ 
to $H_{\Rb}^* (G)$. Replacing $(A,G,\a)$ by $(A \ot C (S^1) 
, G \ts \Rb , \a')$, where $\a'_{g,s} (x \ot f) = \a_g (x) \ot f_s$ and 
$f_s (t) = f(t-s)$, $\fl \, t \in S^1 = \Rb / \Zb$ one extends ${\rm 
ch}_{\tau}$ to a morphism from $K(A) = K_0 (A) \op K_1 (A)$ to 
$H_{\Rb}^* (G)$.

\section*{Special cases} 

$ \ $

(a) Let $(\a_t)_{t \in \Rb}$ be a one parameter group of 
automorphisms of $A$, $\d$ the corresponding derivation, 
then for $U \in A$, invertible and of class $C^{\ify}$ 
one has ${\rm ch}_{\tau} ([U]) = (1/2 \, i \pi) \, \tau (\d (U) U^{-1})$. 
(The non-triviality of this expression was known, cf. 
\cite{4}.)

\smallskip

(b) Let $(\a_{t_1 , t_2})_{t_1 , t_2 \in \Rb}$ be a two parameter group of 
automorphisms of $A$, $\d_1$, $\d_2$ the corresponding 
derivations 
of $A$ and $\ve^1$, $\ve^2$ the canonical basis of
$(\hbox{Lie} \ G)^*$, where $G = \Rb^2$. For every orthogonal projection $e$ 
in $A$ one has ${\rm ch}_{\tau} (e) = \tau (e) + c_1 (e) \, \ve^1 \wdg \ve^2 
\in H_{\Rb}^* (G) = \Lb (\hbox{Lie} \ G)^*$, where $c_1 (e)$ is computed 
for $e$ of class $C^{\ify}$ by the formula
$$
c_1 (e) = \frac{1}{2 \, i \, \pi} \ \tau (e (\d_1 (e) \, \d_2 (e) - \d_2 
(e) \, \d_1 (e))) \, .
$$
In particular $c_1 (e) = 0$ if $e$ is equivalent to an orthogonal projection
fixed by a one parameter subgroup.

Let us now give a concrete example, let $\t \in [0,1] \bsh \Qb$ and 
$A_{\t}$ the $C^*$-algebra generated by two unitaries $U_1$, 
$U_2$ such that $U_1 \, U_2 = \lb \, U_2 \, U_1$, $\lb = \exp (i \, 2 \, 
\pi \, \t)$. The theorem of Pimsner and Voiculescu \cite{3} determines 
$K_0 (A_{\t}) = \Zb^2$ and shows that the unique trace $\tau$ [with $\tau (1) 
= 1$] on $A_{\t}$ defines an isomorphism of $K_0 (A_{\t})$ on 
$\Zb + \t \, \Zb \sbs \Rb$. Moreover the construction of Powers and Rieffel 
\cite{5} exhibits an idempotent $e \in A_{\t}$ with $\tau (e) = \t$~; let 
$\rho$ be the isomorphism of $C(S^1)$, $S^1 = \Rb / \Zb$ in 
$A_{\t}$ which to the function $t \mpo \exp (i \, 2 \, \pi \, t)$ associates 
$U_1$. One has $e_0 = \rho (g) \, U_2 + \rho (f) + (\rho (g) \, U_2)^*$, the 
conditions on $f$, $g \in C (S^1)$ are fulfilled if $f(s) = 1$, 
$\fl \, s \in [1-\t , \t]$, $f(s) = 1 - f (s-\t)$, $\fl \, s \in [\t , 
1]$ and $g(s) = 0$ if $s \in [0,\t]$, $g(s) = (f(s) - f^2 (s))^{1/2}$ if 
$s \in [\t , 1]$.

Let us endow $A_{\t}$ with the action of $\Tb^2$, where $\Tb = \{ z \in 
\Cb , \vert z \vert = 1 \}$ such that $\a_{z_1 , z_2} (U_j) = z_j \, 
U_j$, $j=1,2$ the corresponding derivations fulfill $\d_k (U_k) 
= 2 \, \pi \, i \, U_k$, $\d_j (U_k) = 0$ if $j \ne k$. The algebra 
$A_{\t}^{\ify}$ is the space of sequences with rapid decay  
$(s_{n,m})_{n,m \in \Zb^2}$  the associated element of $A_{\t}$ being 
$\Si \, s_{n,m} \, U_1^n \, U_2^m$. For $f$ and $g$ of class $C^{\ify}$ 
one has $e_0 \in A_{\t}^{\ify}$ and the computation shows that $c_1 (e_0) = -6 
\int g^2 \, f' \, dt = 6 \int_0^1 (\lb - \lb^2) \, d\lb = 1$. One then concludes 
that if $e \in A_{\t}$ is an idempotent such that $\tau (e) = 
\vert p-q \, \t \vert$, one has
$$
c_1 (e) = \pm \, q \, .
$$
This example shows the non-triviality of $c_1$, the integrality 
property of $c_1 \in \Zb$ will be explained later (th.~10), 
let us now concretely realise the finite projective module 
$\Xi_{p,q}^{\ify}$ of dimension $\vert p-q \, \t \vert$ on 
$A_{\t}^{\ify}$.

Let $\Sc (\Rb)$ be the Schwartz space of $\Rb$, we view $\Sc 
(\Rb) \ot \Cb^q$ as the space of sections of the trivial bundle with fiber $\Cb^q$ 
on $\Rb$, so that $\xi (s) \in \Cb^q$, $\fl \, \xi \in \Sc (\Rb) 
\ot \Cb^q$, $s \in \Rb$. Let $W_1$, $W_2$ be two unitaries in 
$\Cb^q$ such that $W_1 \, W_2 = \exp (i \, 2 \, \pi / q) \, W_2 \, W_1$. We view 
 $\Sc (\Rb) \ot \Cb^q$ as a right $A_{\t}^{\ify}$-module letting~:
$$
(\xi \cdot U_1) (s) = W_1 \, \xi (s-\ve) \, , \qquad \fl \, s \in \Rb 
\, , \quad \hbox{where} \ \ve = \frac{p}{q} - \t \, ,
$$
$$
(\xi \cdot U_2) (s) = \exp (i \, 2 \, \pi \, s) \, W_2^p \, \xi (s) \,  
\qquad 
\fl \, s \in \Rb \, .
$$

\medskip

\begin{theorem}
The $A_{\t}^{\ify}$ module thus defined is projective of finite type, of 
dimension $\vert p-q \, \t \vert$, the equalities
$$
(\nb_1 \, \xi) (s) = \frac{d}{ds} \ \xi (s) \quad \hbox{and} \quad 
(\nb_2 \, \xi) (s) = \frac{2 \, \pi \, i \, s}{\ve} \ \xi (s)
$$
define a connection of constant curvature equal to $1/(\t - 
p/q)$.
\end{theorem}

\medskip

One thus obtains another way to compute $c_1$, indeed  
the integral of the curvature is $\vert p-q \, \t \vert / (\t - (p/q)) 
= \pm \, q$.

In this example of $A_{\t}$ the $C^*$-algebra is far  
from trivial (it is not of type I), the differential structure 
coming from the derivations $\d_1$, $\d_2$ is however as regular
as for a compact smooth manifold, in particular~:
\begin{itemize}
\item[1$^{\rm o}$] with $\D = \d_1^2 + \d_2^2$ the operator 
$(1-\D)^{-1}$ from $A_{\t}$ to $A_{\t}$ is a compact operator (in the
usual sense)~;
\item[2$^{\rm o}$] the space $A_{\t}^{\ify}$ is a nuclear space 
(in the sense of Grothendieck).
\end{itemize}

To elaborate on these facts and tie them with the integrality of the 
coefficient $c_1$, let us first remark that (cf. \cite{2}) the crossed product of 
$A_{\t}$ by the above action of $\Tb^2$ is the elementary
$C^*$-algebra  $k$ of compact operators in Hilbert space, and let us then
go back to the general case of a $C^*$ 
dynamical system $(A,G,\a)$, with as a goal the study of elliptic
differential operators of the form (with $G = \Rb^n$) $D = 
{\displaystyle \sum_{\vert \a \vert \leq k}} \ a_j \, \d^j$, where $j = 
(j_1 , \ldots , j_n)$, $a_j \in A^{\ify}$ and where for all $\xi = 
(\xi_1 , \ldots , \xi_n) \in \Rb_n$, $\s (\xi) = i^k \ {\displaystyle 
\sum_{\vert j \vert = k}} \ a_j \, \xi^j$ is {\it invertible} in 
$A$.

\section*{Pseudo differential calculus and $C^*$ dynamical systems}

To simplify let us assume that $G = \Rb^n$, let $B$ be the crosed product 
 $B = A \ts_{\a} \Rb^n$, $A \sbs M(B)$ the canonical isomorphism of 
 $A$ as a subalgebra of the multiplier algebra $M(B)$ of 
$B$ and $s \ra V_s$ the canonical unitary representation of $\Rb^n$ 
on $M(B)$ such that $V_s \, x \, V_s^* = \a_s (x)$, $\fl \, x \in A$.

We endow the involutive algebra $A^{\ify}$ with the family of semi-norms~:
 $p_i (a) = \Vert \d_1^{i_1} \ldots \d_n^{i_n} \, a 
\Vert$, where $i_1 , \ldots , i_n \in \Nb$ and let $\Sc (\Rb^n , 
A^{\ify})$ be the corresponding Schwartz space~: the map $s \ra 
a(s)$ from $\Rb^n$ to $A^{\ify}$ is in $\Sc$ iff for all  
multi-indices $i,j$ the function $p_i ((\partial / \partial s)^j \, 
a(s))$ is of rapid decay.

The $C^*$-algebra $B = A \ts_{\a} \, \Rb^n$ is the norm closure 
of the involutive algebra
$$
\Si = \left\{ \int a(s) \, V_s \, ds \, , \ a \in {\Sc} (\Rb^n , 
A^{\ify}) \right\} \, .
$$

We now construct multipliers of $B$ of the form
$\int a(s) \, V_s \, ds$, where $a$ is a distribution with values 
in $A^{\ify}$ with singular support contained in $\{ 0 \} \sbs \Rb^n$.

\medskip

\begin{definition}
Let $m \in \Zb$, $\Rb_n = (\Rb^n)^{\wdg}$ the dual of $\Rb^n$, and 
$\rho$ a map of class $C^{\ify}$ from $\Rb_n$ to $A^{\ify}$. 
We shall say that $\rho$ is a symbol of order $m$, $\rho \in S^m$ iff~:
\begin{itemize}
\item[$1^{\rm o}$] for all multi-indices $i,j$, there exists $C_{ij} < 
\ify$ such that
$$
p_i \left(\left( \frac{\partial}{\partial \, \xi} \right)^j \, \rho 
(\xi) \right) \leq C_{ij} (1 + \vert \xi \vert)^{m-\vert j \vert} \, ;
$$
\item[$2^{\rm o}$] there exists $\s \in C^{\ify} (\Rb_n \bsh \{ 0 \} , 
A^{\ify})$ such that when $\lb \ra + \ify$ one has $\lb^{-m} \, \rho (\lb 
\, \xi) \ra \s (\xi)$ $[$for the topology of $C^{\ify} (\Rb_n \bsh \{ 0 
\} , A^{\ify})]$.
\end{itemize}
\end{definition}

\medskip

Let $\rho \in S^m$ and $\hat{\rho}$ its Fourier transform in the  
sense of distributions [by hypothesis $\rho \in \Sc' (\Rb_n , 
A^{\ify})$], it is a distribution with values in $A^{\ify}$ 
given by the oscillating integral
$$
\hat{\rho} \, (s) = \int \rho \, (\xi) \, e^{-is \cdot \xi} \, d\xi \, .
$$
Its singular support is contained in $\{ 0 \} \sbs \Rb^n$. Moreover 
$\hat{\rho} \in \Sc (\Rb^n , A^{\ify})$ iff $\rho$ is of order $-\ify$.

\medskip

\begin{proposition}
\begin{itemize}
\item[{\rm (a)}]For all $\rho \in S^m$ the equality $P_{\rho} = \int 
\hat{\rho} (s) \, V_s \, ds$ defines a multiplier $P_{\rho}$ of the  
 involutive algebra $\Si$.
\item[{\rm (b)}] Let $m_1 , m_2 \in \Zb$, $\rho_j \in S^{m_j}$, $j = 
1,2$. There exists $\rho \in S^{m_1 + m_2}$ such that $P_{\rho} = 
P_{\rho_1} \, P_{\rho_2}$.
\item[{\rm (c)}] If $\rho \in S^0$, then $P_{\rho}$ extends as a multiplier 
of $B = A \ts_{\a} \Rb^n$.
\item[{\rm (d)}] The norm closure $\Ec$ of $\{ P_{\rho} , \rho \in 
S^0 \}$ is a sub $C^*$-algebra of $M(B)$.
\item[{\rm (e)}] Let $S_{n-1}$ be the space of half rays $\Rb^+ \, 
\xi$, $\xi \in \Rb_n \bsh [ 0 \}$ and $P \in \Ec$, there exists $\s (P) 
\in A \ot C (S^{n-1})$ such that
$$
\s (P) (\Rb_+ \, \xi) = \lim_{\tau \ra \ify} \hat{\a}_{\tau \xi} (P) \, .
$$
\item[{\rm (f)}] The sequence $0 \longra A \ts_{\a} \, \Rb^n \build 
\longra_{}^{j} \Ec \build \longra_{}^{T} A \ot C (S^{n-1}) \longra 0$ is 
exact.
\end{itemize}
\end{proposition}

\medskip

In (e) we use the simple norm convergence of multipliers 
(viewed as operators in $B$), and $\hat{\a}$ is the dual action. 
 In (f), $j$ is the canonical inclusion of $B$ in 
$M(B)$ and $\s$ the principal symbol as defined in (e).

To the exact sequence (f) corresponds a six terms exact sequence 
involving $K(B)$, $K(\Ec)$ and $K(A \ot C (S^{n-1}))$. In the particular case 
where $G = \Rb^n$ as above, one can explicitly compute the index, i.e. the map from
 $K_1 (A \ot C (S^{n-1}))$ to $K_0 (B)$. 
One has indeed the Thom isomorphism of $\vp_A$, of $K(A)$ on 
$K(B) = K (A \ts_{\a} \, \Rb^n)$, defined by induction on $n$ by 
iterating the isomorphism described in \cite{1}.

Let $0$ be a point of $S^{n-1}$ and $\lb$ the canonical generator 
of $K (S^{n-1} \bsh \{ 0 \})$, $\Psi_A$ the corresponding isomorphism of 
$K(A)$ with $K(A \ot C_0 (S^{n-1} \bsh \{ 0 \}))$.

\medskip

\begin{theorem}
{\rm (index theorem for flows)} One has
$$
{\rm Index} \ P = \vp_A \circ \Psi_A^{-1} (\tau (P)) \, , \qquad \fl 
\, P \in \Ec \, ,
$$
$\s (P)$ invertible.
\end{theorem}

\medskip

On the one hand this  theorem is finer than the index theorem for
foliations with transverse measures in the case of flows, since it computes the 
 $K_0 (C^* (V , \Fc))$-valued index and not only its composition with the trace.
In particular, it shows that the image of $K_0 (C^* (V,\Fc))$ by the trace
is equal to the  image by the  Ruelle-Sullivan cycle $[C]$ of 
the subgroup of $H^* (V,\Rb)$ range of $K(V)$ by the chern character Ch.
Thus, if $H^n (V,\Rb) 
= 0$ this range is zero (cf. \cite{1} for an application in the case 
 $n=1$).

On the other hand, if $\tau$ is a finite $\a$-invariant trace on $A$ and 
$\hat{\tau}$ is the dual trace,  theorem 9 allows to
compute $\hat{\tau} \circ {\rm Ind}$. For $D$ elliptic, one defines 
${\rm ch}_{\tau} (\s_D)$ as in the classical case as the composition of the  
map ${\rm ch}_{\tau}$ defined above with $\Psi_A^{-1}$, 
one then has~:

\medskip

\begin{theorem}
One has $\hat{\tau} \circ {\rm Index} \, D = \langle {\rm ch}_{\tau} (\s_D),
 v \rangle$, where $v = \ve_1 \wdg \ldots \wdg \ve_n$ is 
the unique element of the canonical basis of $\Lb^n \, \Rb^n$.
\end{theorem}

\medskip

Theorem 10 remains valid when one replaces $\Rb^n$ by a  
 Lie group $G$, in particular let $A = A_{\t}$, $G = \Tb^2$ 
acting on $A_{\t}$ as above, then the crossed product $A_{\t} \ts \Tb^2$ is 
isomorphic to the elementary $C^*$-algebra  $k$, the  dual trace
$\hat{\tau}$ being the standard trace on $k$ which shows
the integrality of $\hat{\tau} \circ {\rm Ind}$, as in the case of 
ordinary compact manifolds.

Let us consider $\Sc (\Rb)$ as the right $A_{\t}^{\ify}$-module $\Xi_{(0,1)}$ 
described in theorem 6, then ${\rm End} (\Sc (\Rb))$ is 
the algebra of finite difference operators of the form $(\D F) (s) 
= \sum \, \vp_n (s) \, F(s-n)$, where the $\vp_n$ are periodic smooth functions 
of period $\t$ and where the sequence $\vp_n$ is of rapid decay. 
Thus ${\rm End} (\Sc (\Rb))$ is isomorphic 
to $A_{\t'}^{\ify}$, where $\t' = 1/\t - E (1/\t)$. Taking for instance 
$f$ and $g$ periodic with period 1 fulfilling the above conditions 
 relative to $\t'$, $h(s) = g(s \, \t)$, $k(s) = 1-2 \, f 
(s \, \t)$, the index of the operator
$$
P : (PF) (s) = h(s) \, F' (s-1) + h (s+1) \, F' (s+1) + k(s) \, F' (s) 
+ s \, F(s)
$$
is equal to $1+E(1/\t)$, hence the existence of nonzero solutions of 
equations $PF = 0$, $F \in \Sc (\Rb)$, $F \ne 0$.

\medskip

{\it Remark.} Let $V$ be a compact $n$-dimensional smooth manifold, 
$\Psi$ a diffeomorphism of $V$ and $A$ the crossed-product
$C^*$-algebra of $C(V)$ by the  automorphism 
$\Psi^*$, $\Psi^* \, f = f \circ \Psi$. Let $\Lb V$ be the complex of 
smooth differential forms on $V$. One can define,
 essentially as the crossed-product of $\Lb V$ 
by the action of $\Psi$, a complex $\Om = {\displaystyle \sum_0^{n+1}} 
\ \Om^k$.  The dense involutive subalgebra $\Om^0$ of $A$, 
has properties similar to the complex $\Om = A^{\ify} \ts \Lb 
(\hbox{Lie} \ G)$ discussed above. The differentable structure thus defined
does not in general correspond to derivations of $A$.

The original paper appeared as
{  A. Connes}, {\it  $C^*$ alg\`ebres et
g\'eom\'etrie differentielle}.  C.R. Acad. Sci. Paris,
Ser.~A-B , 290, 1980.
The above references appeared respectively as,
[1] {\it Adv. in Math.} {\bf 
39} (1981)  31-55.
[2]  {\it J. Operator theory} {\bf 
3} (1980)  237-269.
[3]  {\it J. Operator Theory} {\bf 
4} (1980), 93-118.
[4]  {\it Comm. Math. Phys.}  {\bf 
74} (1980), 281-295.
[5]   {\it Short 
comm.} I.C.M., 1978.
[6]  {\it Adv. Math. Suppl. Stud.} {\bf 
78} Academic Press (1983) 155-163.

\end{document}